\documentclass[a4paper,12pt]{article}

\def\empha{\em}

\def\bemphas{}
\def\pemphas{}
\def\scsc{\sc}
\newtheorem{thm}{Theorem}[section]

\newtheorem{defi}[thm]{Definition}

\long
\def\MSC#1\EndMSC{\def\arg{#1}\ifx\arg\empty\relax\else
      {\par\narrower\noindent%
      2000 Mathematics Subject Classification. #1\par}\fi}

\long
\def\KEY#1\EndKEY{\def\arg{#1}\ifx\arg\empty\relax\else
    {\par\narrower\noindent%
      Keywords and Phrases: #1\par}\fi}

\title
{Classical phase space singularities and quantization}

\author{Johannes Huebschmann \footnote{Support by the Deutsche
Forschungsgemeinschaft in the framework of a
Mercator-professorship is gratefully
acknowledged}\\
Universit\'e des
Sciences et Technologies de Lille\\ UFR de Math\'ematiques, CNRS-UMR 8524
\\ 59655 VILLENEUVE D'ASCQ, C\'edex, France
\\ and
\\ Institute
for Theoretical Physics\\ Universit\"at Leipzig\\
04109 LEIPZIG, Germany \\
{Johannes.Huebschmann@math.univ-lille1.fr} }

\usepackage{amssymb}
\usepackage[square,authoryear]{natbib}
\usepackage{amsmath}
\usepackage{amscd}

\numberwithin{equation}{section}

\begin{document}
\setcounter{page}{1}

\maketitle

\begin{abstract}
Simple classical mechanical systems and solution spaces of
classical field theories involve singularities. In certain
situations these singularities can be understood in terms of the
stratified K\"ahler spaces. We give an overview of a research
program whose aim is to develop a holomorphic quantization
procedure on stratified K\"ahler spaces.
\end{abstract}

\MSC Primary:  17B63 17B65 17B66 17B81 53D17 53D20 53D50 70H45
81S10; Secondary: 14L24 14L30 32C20 32Q15 32S05 32S60
 \EndMSC

\KEY
Stratified symplectic space,
complex analytic space,
complex analytic stratified K\"ahler space,
reduction and quantization,
holomorphic quantization,
quantization on a stratified K\"ahler space,
constrained system,
invariant theory,
hermitian Lie algebra,
correspondence principle,
Lie-Rinehart algebra,
adjoint quotient
\EndKEY

\section{Quantum theory and classical singularities}

According to {\scsc Dirac}, the {\em correspondence\/} between a
classical theory and its quantum counterpart should be based on an
analogy between their mathematical structures.
An interesting
issue is then that of the role of singularities in quantum
problems. Singularities are known to arise in classical phase
spaces. For example, in the hamiltonian picture of a theory,
reduction modulo gauge symmetries leads in general to
singularities on the classical level. Thus we are running into the
question what the significance of singularities on the quantum
side could be. Can we ignore them, or is there a quantum structure
which has the classical singularities as its shadow?
 As far as know, one of the first
papers in this topic is that of {\scsc Emmrich and R\"omer\/}
\cite{emmroeme}. This paper indicates that wave functions may
\lq\lq congregate\rq\rq\ near a {\em singular\/} point, which goes
counter to the sometimes quoted statement that {\em singular
points in a quantum problem are a set of measure zero so cannot
possibly be important.\/} In a similar vein, {\scsc Asorey et
al\/} observed that vacuum nodes correspond to the chiral gauge
orbits of reducible gauge fields with non-trivial magnetic
monopole components \cite {asfalolu}. It is also noteworthy that
in classical mechanics and
in classical field theories  singularities in the solution spaces
are the {\em rule rather than the exception\/}.
This is in particular true for Yang-Mills theories and
for Einstein's gravitational theory where
singularities occur even at some of
the most interesting and physically relevant solutions, namely at the
symmetric ones.
It is still not
understood what role these singularities might have in quantum
gravity. See, for example, {\scsc Arms, Marsden and Moncrief\/}
\cite{armamonc}, \cite{armamotw} and the literature there.

\section{An example of a classical phase space singularity}
\label{example}
In $\mathbb R^3$ with coordinates $x,y,r$,
consider the semicone $N$
given by the equation $x^2 + y^2 = r^2$ and the inequality $r \geq 0$.
We refer to this semicone as the {\em exotic\/} plane with a single
vertex. The semicone $N$ is the classical reduced
{\em phase space\/} of a single particle
moving in ordinary affine space of
dimension $\geq 2$ with angular momentum zero.
This claim will actually be justified in Section \ref{illustration}
below. The reduced Poisson algebra
$(C^{\infty}N,\{\,\cdot\,,\,\cdot\,\})$ may be described in the following
fashion: Let $x$ and $y$ be the ordinary coordinate functions in
the plane, and consider the algebra $C^{\infty}N$ of smooth
functions in the variables $x,y,r$ subject to the relation $x^2 +
y^2 = r^2$. Define the Poisson bracket $\{\,\cdot\,,\,\cdot\,\}$ on
this algebra by
\[
\{x,y\} = 2r,\  \{x,r\} = 2y,\ \{y,r\} = -2x,
\]
and endow $N$ with the complex structure having $z=x+iy$ as
holomorphic coordinate. The Poisson bracket is then {\em defined
at the vertex\/} we well, away from the vertex the Poisson
structure is an ordinary {\em symplectic\/} Poisson structure, and
the complex structure does {\em not\/} \lq\lq see\rq\rq\ the
vertex. At the vertex, the radius function $r$ is {\em not\/} a
smooth function of the variables $x$ and $y$. Thus the vertex is a
singular point for the Poisson structure whereas it is {\em not\/}
a singular point for the complex analytic structure. The Poisson
and complex analytic structures
 combine to a \lq\lq stratified K\"ahler structure\rq\rq.
We will explain shortly what this means.

\section{Stratified K\"ahler spaces}

In the presence of singularities, restricting quantization to a
smooth open dense stratum, sometimes referred to as
\lq\lq top stratum\rq\rq, can result
in a loss of information and may in fact lead to {\it
inconsistent\/} results.
To develop a satisfactory notion of K\"ahler quantization
in the presence of singularities,
on the classical level, we isolated
a notion of \lq\lq K\"ahler space
with singularities\rq\rq;
we refer to such a space as
a {\em stratified K\"ahler
space\/}.
Ordinary {\em K\"ahler quantization\/}
may then be extended to
a {\em quantization scheme over stratified K\"ahler spaces\/}.

We will now explain the concept of a
\emph{complex analytic stratified K\"ahler
space\/}. In \cite{kaehler} we introduced a notion of stratified
K\"ahler space which includes that of complex analytic stratified
K\"ahler space. For the sake of clarity, we maintain the distinction between
a (possibly more general) stratified K\"ahler space and a complex
analytic one. We do not know whether the two notions
really differ. For the present paper, the notion of
complex analytic stratified K\"ahler
space suffices.

Let $N$ be a stratified space.
We recall first that
a {\em stratified symplectic structure\/} on $N$ is a Poisson algebra
$(C^{\infty}N,\{\,\cdot\,,\,\cdot\,\})$ of continuous
functions on $N$ having the following properties:
\\
{\rm (1)}
Each stratum is an ordinary smooth manifold,
and the restriction map from
$C^{\infty}N$ to the algebra of continuous functions on that stratum
goes into the algebra of ordinary smooth functions on the stratum.
\\
{\rm (2)}
Each stratum carries a symplectic structure, and the
restriction map from $C^{\infty}N$ to the
algebra of smooth functions on the stratum
is a morphism of Poisson algebras, where the stratum
is endowed with the ordinary smooth symplectic Poisson structure.
\\
A \emph{stratified symplectic space\/}
is a stratified space together with a stratified symplectic structure.
The functions in $C^{\infty}N$ are not necessarily
ordinary smooth functions.
In the special case
where restriction of the functions in $C^{\infty}N$
to any stratum yields
the compactly supported functions on that stratum,  the
Poisson algebra
$(C^{\infty}N,\{\,\cdot\,,\,\cdot\,\})$
actually induces the symplectic Poisson structure on each stratum.

Next we recall that a {\em complex analytic space\/}
(in the sense of {\scsc Grauert\/}) is a topological space $X$, together
with a sheaf of rings $\mathcal O_X$, having the following property:
The space $X$ can be covered by open sets $Y$, each of which
embeds into the open polydisc
$U=\{\mathbf z=(z_1,\dots,z_n);|\mathbf z|<1\}$ in some $\mathbb C^n$
(the dimension $n$ may vary as $U$ varies)
as the zero set of a finite system of holomorphic functions
$f_1,\dots,f_q$ defined on $U$,
such that the
restriction $\mathcal O_Y$ of the sheaf $\mathcal O_X$ to $Y$
is isomorphic as a sheaf to
the quotient sheaf $\mathcal O_U\big/(f_1,\dots,f_q)$;
here $\mathcal O_U$ is the sheaf of germs of holomorphic functions
on $U$. The sheaf $\mathcal O_X$ is then referred to as the
{\em sheaf of holomorphic functions on \/} $X$.
See \cite{gunnross} for a development of the general theory of complex
analytic spaces.

\begin{defi}
 A complex analytic {\em stratified K\"ahler
space\/} consists of a stratified space $N$, together with
\\
{\rm (i)} a stratified symplectic structure
$(C^{\infty}N,\{\,\cdot\,,\,\cdot\,\})$ having the given
stratification of $N$ as its underlying stratification, and with
\\
{\rm (ii)} a complex analytic structure on $N$ which is compatible
with the stratified symplectic structure.
\end{defi}

\noindent The two structures being \emph{compatible\/} means the
following:
\\
(i) Each stratum is a complex analytic subspace, and the complex
analytic structure, restricted to the stratum, turns that stratum
into an ordinary complex manifold; in particular, the
stratification of $N$ is a refinement of the complex analytic
stratification.
\\
(ii) For each point $q$ of $N$ and each holomorphic function $f$
defined on an open neighborhood $U$ of $q$, there is an open
neighborhood $V$ of $q$ with $V \subset U$ such that, on $V$, $f$
is the restriction of a function in $C^{\infty}(N,\mathbb C) =
C^{\infty}(N) \otimes \mathbb C$.
\\
(iii) On each stratum, the symplectic structure  combines with the
complex analytic structure to a K\"ahler structure.

\noindent {\scsc Example 1\/}: The {\em exotic plane\/}, endowed with
the structure explained in Section \ref{example} above, is a
stratified K\"ahler space. Here the radius function $r$ is {\em not\/} an
ordinary smooth function of the variables $x$ and $y$. Thus the stratified
symplectic structure cannot be given in terms of ordinary smooth functions
of the variables $x$ and $y$.

This example generalizes to an entire class of examples:
The {\em closure of a holomorphic nilpotent orbit\/}
(in a hermitian Lie algebra)
inherits a complex analytic stratified K\"ahler structure
\cite{kaehler}.
Angular momentum zero reduced spaces are special cases thereof;
see Section \ref{illustration} below for details.

{\em Projectivization\/} of  the closure of a holomorphic
 nilpotent orbits yields what we call an {\em exotic projective variety\/}.
This includes  complex
quadrics, {\scsc Severi\/} and {\scsc Scorza\/} varieties and their
{\em secant\/} varieties \cite{kaehler}, \cite{scorza}.
In physics, spaces of this kind arise as reduced classical phase spaces for
systems of harmonic oscillators with zero angular momentum
and constant energy. We shall explain some of the details in
Section \ref{illustration} below.

\noindent {\scsc Example 2\/}: A moduli space of semistable
holomorphic vector bundles or, more generally, a moduli space of
semistable principal bundles on a non-singular complex projective
curve carries a complex analytic stratified K\"ahler structure
\cite{kaehler}. As a space, that is, the complex analytic
structure being ignored, such a moduli space arises as the moduli
space of homomorphisms or more generally twisted homomorphisms
from the fundamental group of a (real) surface to a compact
connected Lie group as well. In conformal field theory, moduli
spaces of this kind occur as spaces of {\em conformal blocks\/}.
The construction of the moduli spaces as complex projective
varieties goes back to \cite{narasesh} and \cite{seshaone}; see
\cite{seshaboo} for an exposition of the general theory. Atiyah
and Bott \cite{atibottw} initiated another approach to the study
of these moduli spaces by identifying them with moduli spaces of
projectively flat constant central curvature connections on
principal bundles over Riemann surfaces, which they analyzed by
methods of gauge theory. In particular, by applying the method of
symplectic reduction to the action of the infinite-dimensional
group of gauge transformations on the infinite-dimensional
symplectic manifold of all connections on a principal bundle, they
showed that an invariant inner product on the Lie algebra of the
Lie group in question induces a natural symplectic structure on a
certain smooth open and dense stratum which, together with the
complex analytic structure determined by the complex structure on
the Riemann surface (complex curve), turns that stratum into an
ordinary K\"ahler manifold. In certain cases, the open and dense
stratum is the entire space, which is then a compact K\"ahler
manifold. This infinite-dimensional approach to moduli spaces of
the kind under discussion has  roots in quantum field theory.
Thereafter a finite-dimensional construction of such a moduli
space as a symplectic quotient arising from an ordinary
finite-dimensional Hamiltonian $G$-space for a compact Lie group
$G$ was developed; see  \cite{modus}, \cite{oberwork} and the
literature there. This construction exhibits the moduli space as a
\emph{compact\/} stratified symplectic space. The stratified
symplectic structure, in turn, combines with the complex analytic
structure determined by the complex structure on the Riemann
surface to a complex analytic stratified K\"ahler structure. This
structure includes the K\"ahler manifold structure on the open and
dense stratum; indeed it compactifies this K\"ahler manifold to a
\emph{compact\/} complex analytic stratified K\"ahler space.

An important special case is that of the moduli space of
semistable rank 2 degree zero vector bundles with trivial
determinant on a curve of genus 2. As a space, this moduli space
is a copy of ordinary complex projective 3-space, but the
stratified symplectic structure involves more functions than just
ordinary smooth functions. In this moduli space, as a complex
analytic subspace, the complement of the space of stable vector
bundles is a {\em Kummer surface\/}, associated with the Jacobian
of the curve. Endowed with the additional structure of a complex
analytic stratified K\"ahler space, the Kummer surface is another
example of an \emph{exotic\/} projective variety.
 See \cite{locpois}--\cite{oberwork}, \cite{kaehler}, and the literature
there.

Any ordinary K\"ahler manifold is
plainly a complex analytic
stratified K\"ahler space.
This kind of example generalizes in the following fashion:
For a Lie group $K$, we will
denote its Lie algebra by $\mathfrak k$ and the dual
thereof by
$\mathfrak k^*$.
The next result says that, roughly speaking,
K\"ahler reduction, applied to an ordinary K\"ahler manifold,
yields  a complex analytic
stratified K\"ahler structure on the reduced space.

\begin{thm}[\cite{kaehler}]
\label{kaehler1}
Let $N$ be a K\"ahler manifold, acted upon holomorphically
by a complex Lie group $G$ such that the action, restricted to
a compact real form $K$ of $G$, preserves the
K\"ahler structure and is hamiltonian, with momentum mapping
$\mu \colon N \to \mathfrak k^*$. Then the reduced space
$N_0 = \mu^{-1}(0)\big/K$ inherits a complex analytic
stratified K\"ahler structure.
\end{thm}

For intelligibility, we explain briefly how the structure on the
reduced space $N_0$ arises. Details may be found in
\cite{kaehler}: Consider the algebra $C^{\infty}(N)^K$ of smooth
$K$-invariant functions on $N$, and let $I^K$ be the ideal of
functions in $C^{\infty}(N)^K$ that vanish on the zero locus
$\mu^{-1}(0)$. Define $C^{\infty}(N_0)$ to be the quotient algebra
$C^{\infty}(N)^K\big/I^K$. This is an algebra of continuous
functions on $N_0$ in an obvious fashion. The ordinary smooth
symplectic Poisson structure $\{\,\cdot\,,\,\cdot\,\}$ on
$C^{\infty}(N)$ is $K$-invariant and hence induces a Poisson
structure on the algebra $C^{\infty}(N)^K$ of smooth $K$-invariant
functions on $N$. Furthermore, {\scsc Noether\/}'s theorem entails
that the ideal $I^K$ is a Poisson ideal, that is to say, given $f
\in C^{\infty}(N_0)^K$ and $h\in I^K$, the function $\{f,h\}$ is
in $I^K$ as well. Consequently the Poisson bracket
$\{\,\cdot\,,\,\cdot\,\}$ descends to a Poisson bracket
$\{\,\cdot\,,\,\cdot\,\}_0$ on $C^{\infty}(N_0)$. Relative to the
orbit type stratification, the Poisson algebra
$(C^{\infty}N_0,\{\,\cdot\,,\,\cdot\,\}_0)$ turns $N_0$ into a
stratified symplectic space.

The inclusion of $\mu^{-1}(0)$ into $N$ passes to a
homeomorphism from
$N_0$ onto the
categorical $G$-quotient
$N\big/\big/ G$ of $N$ in the category of
complex analytic varieties.
The stratified symplectic structure combines with the
complex analytic structure on
$N\big/\big/ G$ to a stratified K\"ahler structure.
When $N$ is complex algebraic, the complex algebraic $G$-quotient
coincides with the complex analytic $G$-quotient.

Thus we see that, in view of Theorem \ref{kaehler1}, examples of
stratified K\"ahler spaces abound.

\noindent {\scsc Example 3\/}: Adjoint quotients of
complex reductive Lie groups; see \cite{adjoint} for details:
Let $G$ be a compact Lie group,
and let $G^{\mathbb C}$ be its complexification;
any complex reductive Lie group is of this kind.
Let $\mathfrak g$ be the Lie algebra of $G$.
Recall that the polar map
from $G \times \mathfrak g$ to $G^{\mathbb C}$, which
is given by the assignment to
$(x,Y) \in G \times \mathfrak g$ of $x\, \mathrm{exp}(iY) \in G^{\mathbb C}$,
is a $G$-biinvariant diffeomorphism.
We endow the Lie algebra $\mathfrak g$  with an
invariant (positive definite) inner product;
by means of this inner product,
we identify $\mathfrak g$ with its dual $\mathfrak g^*$ and,
furthermore,
the total space $\mathrm TG$ of the tangent bundle of $G$  with the total
space $\mathrm T^*G$ of the cotangent bundle of $G$.
The composite
\[
\mathrm T^*G \longrightarrow G \times \mathfrak g \longrightarrow
G^{\mathbb C}
\]
of left translation with the polar map
is a diffeomorphism, and it may be shown that the resulting complex structure
on $\mathrm T^*G$ combines with the  cotangent bundle symplectic structure
to a K\"ahler structure.
Moreover, the action of
$G^{\mathbb C}$ on itself
 by conjugation is holomorphic, and the restriction
of the action to $G$ is Hamiltonian, with momentum mapping
from $G^{\mathbb C}$ to $\mathfrak g^*$ which, viewed as a map on the
isomorphe $G\times \mathfrak g$ and with values in $\mathfrak g$,
amounts to the map
\[
\mu \colon G\times \mathfrak g \longrightarrow \mathfrak g,
\quad
\mu(x,Y) = \mathrm{Ad}_xY - Y.
\]
By
Theorem \ref{kaehler1}, the reduced space $(\mathrm T^*G)_0$ inherits a
stratified K\"ahler space structure.
In physics,  a space of the kind $(\mathrm T^*G)_0$
is the {\em building block\/} for certain {\it
lattice gauge\/} theories.

Consider the special case where
$G=\mathrm{SU}(n)$, so that
$G^{\mathbb C}=\mathrm{SL}(n,\mathbb C)$.
We claim that, complex analytically, the reduced space $(\mathrm T^*G)_0$
amounts to a copy $\mathbb C^{n-1}$
of complex affine $(n-1)$-dimensional space.
Indeed, the characters
$\sigma_1, \dots,
\sigma_{n-1}$ of the fundamental representations
of $\mathrm{SL}(n,\mathbb C)$
yield a holomorphic map
\[
 (\sigma_1,\dots,\sigma_{n-1})\colon \mathrm{SL}(n,\mathbb C)
 \longrightarrow
\mathbb C^{n-1},
\]
referred to in the literature as {\scsc Steinberg\/} map,
such that the induced map
from $\mathrm{SL}(n,\mathbb C)\big/\big/\mathrm{SL}(n,\mathbb C)$
to $\mathbb C^{n-1}$ is an isomorphism of complex algebraic
and hence of complex analytic spaces.
Hence, as a complex analytic space,
the reduced space
$(\mathrm T^* \mathrm{SL}(n,\mathbb C))_0$
comes down  to
a copy $\mathbb C^{n-1}$
of complex affine $(n-1)$-dimensional space.

Here is another way to understand the situation:
A maximal complex torus $T^{\mathbb C}$
in $\mathrm{SL}(n,\mathbb C)$
is given by the complex diagonal matrices in
$\mathrm{SL}(n,\mathbb C)$, that is,
 by the complex diagonal $(n \times n)$-matrices having determinant 1.
Realize the torus $T^{\mathbb C}$ as the subspace of $(\mathbb
C^*)^n$ which consists of all $(z_1,\dots,z_n)$ in $(\mathbb
C^*)^n$ such that $z_1 \dots z_n = 1$; then the characters
$\sigma_1, \dots, \sigma_{n-1}$, restricted to $T^{\mathbb C}$,
are the elementary symmetric functions in the variables
$z_1,\dots,z_n$. The Steinberg map restricts to the holomorphic
map
\[
 (\sigma_1,\dots,\sigma_{n-1})\colon (\mathbb C^*)^{n-1}
 \longrightarrow
\mathbb C^{n-1}
\]
given by the assignment to
$\mathbf z=(z_1,\dots,z_n)\in T^{\mathbb C}$ of
$(\sigma_1(\mathbf z),\dots,\sigma_{n-1}(\mathbf z))$.
Since, for a general
complex Lie group of the kind $G^{\mathbb C}$,
 every conjugacy class has in its closure a unique
semisimple conjugacy class and since
semisimple conjugacy classes are parametrized
by orbits of the action of the Weyl group on the maximal complex
torus $T^{\mathbb C}$,
we see that the restriction of the
Steinberg map to the maximal torus
already realizes  the complex analytic quotient
as the space of orbits relative to the action of
the Weyl group on  $T^{\mathbb C}$, which is here the
symmetric group $S_n$ on $n$ letters.

As a {\em stratified K\"ahler space\/}, the adjoint quotient is
considerably more complicated.
We will restrict attention to the
even more special case where $n=2$, so that $G=\mathrm{SU}(2)$
and
$G^{\mathbb C}=\mathrm{SL}(2,\mathbb C)$.
Then
\[
T= \left\{ \left[\begin{array}{cccc} \zeta & 0
\\
0 &  \zeta^{-1}
\end{array}
\right], |\zeta|=1\right\}
\]
is a maximal torus in $G$, and
\[
\mathbb C^* \cong T^{\mathbb C}=
\left\{\left[\begin{array}{cccc} \zeta & 0
\\
0 &  \zeta^{-1}
\end{array}
\right],  \zeta\ne 0\right\}
\]
is a maximal torus in $G^{\mathbb C}=\mathrm{SL}(2,\mathbb C)$. In
view of the above discussion, complex analytically, the adjoint
quotient $G^{\mathbb C}\big/\big/G^{\mathbb C}$ amounts to the
space $T^{\mathbb C}\big /(\mathbb Z\big/2) \cong \mathbb C$ of
orbits relative to the action of the Weyl group $S_2$ on $\mathbb
C^* \cong T^{\mathbb C}$, and this orbit space is realized as the
target of the holomorphic map
\[
f \colon \mathbb C^* \longrightarrow \mathbb C,\ f(z) = z+ z^{-1}.
\]
Thus $Z = z+ z^{-1}$ may be taken as a holomorphic coordinate on the
adjoint quotient.

On the other hand, the real structure is encapsulated by the \emph
{real stratified symplectic Poisson structure\/}
$(C^{\infty}(\mathrm T^*G)_0,\{\,\cdot\ \, , \, \cdot \,\})$
which, in turn, involves the three functions
\[
X = x + \frac{x}{r^2}, \ Y = y - \frac{y}{r^2},\ \tau =
\frac{y^2}{r^2},
\]
where $z=x+iy$, $Z=X+iY$, and $x^2 + y ^2 = r^2$. The resulting
smooth structure $C^{\infty}(\mathrm T^*G)_0$ on the quotient
\[
G^{\mathbb C}\big/\big/G^{\mathbb C} \cong (\mathrm T^*G)_0 \cong
\mathbb C
\]
is the algebra of smooth functions in the variables $X$, $Y$,
$\tau$, subject to the relation
\[
Y^2 = (X^2 + Y^2 + 4 (\tau -1)) \tau.
\]
Moreover, the Poisson bracket $\{\,\cdot\,,\,\cdot\,\}$ on
$C^{\infty}(\mathrm T^*G)_0$ is determined by
\[
\{X,Y\} = X^2 + Y ^2 + 4(2 \tau -1) .
\]
The resulting complex analytic stratified K\"ahler structure is
{\em singular\/} at the points  $-2$ and $2$, that is, the Poisson
structure vanishes at these points; furthermore, at these two
points, the function $\tau$ is {\em not\/} an ordinary smooth
function of the variables $X$ and $Y$. Away from these two points,
the Poisson structure is symplectic. We refer to this adjoint
quotient as the \emph{exotic plane with two vertices\/}. It is
perhaps worthwhile noting that this complex analytic stratified
K\"ahler space also arises as the reduced classical phase space of
a spherical pendulum constrained to move with angular momentum
zero, so that it moves in a plane. See \cite{adjoint} for details.

\section{Correspondence principle and Lie-Rinehart algebras}

To make sense of the \noindent{\em correspondence principle\/} in
certain {\em singular\/} situations, one needs a tool which, for the
stratified symplectic
Poisson algebra  on a stratified symplectic space,
serves as a {\em replacement\/} for
the tangent bundle of a smooth symplectic manifold. This
replacement is provided for by an appropriate {\em Lie-Rinehart
algebra\/}. This Lie-Rinehart algebra yields in particular a
satisfactory generalization of the Lie algebra of smooth vector fields in the
smooth case. This enables us to put {\em flesh on the bones of
Dirac's correspondence principle in certain singular
situations\/}.

 A {\em Lie-Rinehart algebra\/} consists of a commutative
algebra and a Lie algebra with additional structure which
generalizes the mutual structure of interaction between the
algebra of smooth functions and the Lie algebra of smooth vector
fields on a smooth manifold. More precisely:

\begin{defi}
  A {\em Lie-Rinehart\/} algebra
consists of a commutative algebra $A$ and a Lie-algebra $L$ such
that $L$ acts on $A$ by derivations and that $L$ has an $A$-module
structure, and these are required to satisfy
\[
\begin{aligned} {} [\alpha,a\beta]  &= \alpha(a)\beta + a [\alpha,\beta],
\\
(a\alpha)(b) &= a (\alpha(b)),
\end{aligned}
\]
where $a,b \in A$ and $\alpha, \beta \in L$.
\end{defi}

We will now explain briefly the Lie-Rinehart algebra associated
with a Poisson algebra; more details may be found in
\cite{poiscoho}, \cite{souriau}, and \cite{lradq}. Thus, let
$(A,\{\,\cdot\,,\,\cdot\,\})$ be a Poisson algebra. Let $D_A$ be
the $A$-module of formal differentials of $A$ the elements of
which we write as $du$, for $u \in A$. For $u,v \in A$, the
association
\[
(du,dv)\longrightarrow \pi (du,dv) = \{u,v\}
\]
yields an $A$-valued $A$-bilinear skew-symmetric 2-form $\pi=
\pi_{\{\,\cdot\,,\,\cdot\,\}}$ on $D_A$, referred to as
the {\em Poisson\/} 2-{\em form\/} associated with
the Poisson structure $\{\,\cdot\,,\,\cdot\,\}$.
The adjoint
\[
\pi^{\sharp} \colon D_A \longrightarrow \mathrm{Der}(A) =
\mathrm{Hom}_A(D_A,A)
\]
of $\pi$ is a morphism of $A$-modules, and the formula
\[
[a du,b dv] = a\{u,b\} dv + b\{a,v\} du + ab d\{u,v\}
\]
yields a Lie bracket $[\,\cdot\,,\,\cdot\,]$ on $D_A$.

\begin{thm}[\cite{poiscoho}] The $A$-module structure on $D_A$, the bracket
$[\cdot,\cdot]$, and the morphism
 $\pi^{\sharp}$ of $A$-modules turn the pair $(A,D_A)$
into a Lie-Rinehart algebra.
\end{thm}

We will write the resulting
Lie-Rinehart algebra as
$(A,D_{\{\,\cdot\,,\,\cdot\,\}})$.
For intelligibility we recall that,
given a Lie-Rinehart algebra $(A,L)$,
the Lie algebra $L$ together with the additional
$A$-module structure
on $L$ and $L$-module structure on $A$
is referred to as an $(\mathbb R,A)$-{\em Lie algebra\/}.
Thus $D_{\{\,\cdot\,,\,\cdot\,\}}$ is an $(\mathbb R,A)$-Lie algebra.

When the Poisson algebra $A$ is the algebra of smooth functions $C^{\infty}(N)$
on a symplectic manifold $N$,
endowed with the ordinary symplectic Poisson structure,
the $A$-dual $\mathrm{Der}(A) =\mathrm{Hom}_A(D_A,A)$
of $D_A$ amounts to the $A$-module $\mathrm{Vect}(N)$ of smooth vector fields,
and
\begin{equation}
(\pi^{\sharp},\mathrm{Id})\colon (D_A,A) \longrightarrow
(\mathrm{Vect}(N),C^{\infty}(N))
\label{morphism}
\end{equation}
is a morphism of Lie-Rinehart algebras,
where $(\mathrm{Vect}(N),C^{\infty}(N))$
carries its ordinary Lie-Rinehart structure.
The $A$-module morphism
$\pi^{\sharp}$ is plainly surjective, and the kernel
consists of those formal differentials which
\lq\lq vanish at each point of\rq\rq\ $N$.

We return to our general Poisson algebra $(A,\{\,\cdot\,,\,\cdot\,\})$.
The Poisson 2-form $\pi_{\{\,\cdot\,,\,\cdot\,\}}$
determines an {\em extension\/}
\begin{equation}
0
\longrightarrow
A
\longrightarrow
\overline L_{\{\,\cdot\,,\,\cdot\,\}}
\longrightarrow
D_{\{\,\cdot\,,\,\cdot\,\}}
\longrightarrow
0
\label{extension}
\end{equation}
of $(\mathbb R,A)$-Lie algebras which is central
as an extension of ordinary Lie algebras;
in particular, on the kernel $A$, the Lie bracket is trivial.
Moreover, as $A$-modules,
\begin{equation}
\overline L_{\{\,\cdot\,,\,\cdot\,\}} = A \oplus D_{\{\,\cdot\,,\,\cdot\,\}},
\label{directsum}
\end{equation}
and the Lie bracket on $\overline L_{\{\,\cdot\,,\,\cdot\,\}}$
is given by
\begin{equation}
[(a,du),(b,dv)] =
\left(
\{u,b\}+ \{a,v\} - \{u,v\}, d\{u,v\}
\right) ,\quad
a,b,u,v \in A.
\label{liebracket}
\end{equation}
Here we have written \lq\lq $\overline L$\rq\rq\
rather than simply $L$ to indicate that
the extension \eqref{extension} represents the {\em negative\/} of the class of
$\pi_{\{\,\cdot\,,\,\cdot\,\}}$
in Poisson cohomology
$\mathrm H_{\mathrm{Poisson}}^2(A,A)$, cf. \cite{poiscoho}.
When $(A,\{\,\cdot\,,\,\cdot\,\})$ is the smooth symplectic Poisson algebra
of an ordinary smooth symplectic manifold,
(perhaps) up to sign, the class of $\pi_{\{\,\cdot\,,\,\cdot\,\}}$
comes essentially
down to the cohomology class represented by the symplectic structure.

The following concept has been introduced in \cite{souriau}.

\begin{defi}
\label{prequantummodule} Given an
$(A\otimes \mathbb C)$-module
$M$,
we refer to
an
$(A,\overline L_{\{\,\cdot\,,\,\cdot\,\}})$-module
structure
\begin{equation}
\chi
\colon
\overline L_{\{\,\cdot\,,\,\cdot\,\}}
\longrightarrow
\mathrm{End}_{\mathbb R}(M)
\label{prequantum}
\end{equation}
on $M$
as a
{\em prequantum module structure for\/}
$(A,{\{\,\cdot\,,\,\cdot\,\}})$
provided \\
{\rm (i)} the values of $\chi$ lie in
$\mathrm{End}_{\mathbb C}(M)$,
that is to say,
for $a \in A$ and $\alpha \in
D_{\{\,\cdot\,,\,\cdot\,\}}$,
the operators $\chi(a,\alpha)$ are complex linear
transformations,
and \\
{\rm (ii)}
for every $a\in A$, with reference to the decomposition \eqref{directsum},
we have
\begin{equation}
\chi(a,0) = i\,a\,\mathrm{Id}_M.
\label{complex}
\end{equation}
A pair $(M,\chi)$
consisting of
an $(A\otimes \mathbb C)$-module $M$ and a prequantum
module structure
will henceforth be referred to as a {\em prequantum module\/}
(for $(A,\{\,\cdot\,,\,\cdot\,\})$.
\end{defi}

{\em Prequantization\/} now proceeds
in the following fashion,
cf. \cite{poiscoho}:
The assignment to $a \in A$ of
$(a,da) \in
\overline L_{\{\,\cdot\,,\,\cdot\,\}}$
yields a morphism $\iota$ of real Lie algebras
from
$A$ to
$\overline L_{\{\,\cdot\,,\,\cdot\,\}}$;
thus, for any prequantum module $(M,\chi)$,
the composite of $\iota$ with $-i \chi$
is a representation
$a \mapsto \widehat a$
of the $A$ underlying real Lie algebra
having $M$, viewed as a complex vector space,
as its representation space;
this is a representation by $\mathbb C$-linear operators
so that any constant acts by multiplication,
that is,
for any real number $r$,
viewed as a member of $A$,
\begin{equation}
\widehat r = r \,\mathrm{Id}
\label{constants}
\end{equation}
and so that, for $a,b \in A$,
\begin{equation}
\widehat {\{a,b\}} = i\,[\widehat a,\widehat b]
\qquad
\text{(the Dirac condition).}
\label{dirac}
\end{equation}
More explicitly, these operators are given by the formula
\begin{equation}
\widehat a (x) = \frac 1 i \chi(0,da) (x) + ax,
\quad
a \in A,\ x \in M.
\label{prequanti}
\end{equation}
In this fashion, prequantization, that is to say,
the first step in the realization of
the correspondence principle in one direction,
can be made precise in certain singular situations.

When $(A,\{\,\cdot\,,\,\cdot\,\})$ is the Poisson algebra of
smooth functions on an ordinary smooth sympletic manifold,
this prequantization factors through the morphism
\eqref{morphism} of Lie-Rinehart algebras in such a way that,
on the target, the construction comes down to the ordinary prequantization
construction.

{\scsc Remark.\/} In the physics
literature, Lie-Rinehart algebras were explored in a paper by
{\scsc Kastler and Stora\/} under the name {\em Lie-Cartan
pairs\/} \cite{kasstora}.

\section{Quantization on stratified K\"ahler spaces}

In the paper \cite{qr} we have shown that
the {\em holomorphic\/} quantization scheme may be extended
to complex analytic stratified K\"ahler spaces.
We recall the main steps:

\noindent 1) The notion of ordinary K\"ahler polarization
generalizes to that of {\em stratified K\"ahler  polarization\/}.
This concept is defined in terms of the Lie-Rinehart
algebra associated with the stratified symplectic Poisson
structure; it
 specifies {\em polarizations on the
strata\/} and, moreover, encapsulates the {\em mutual positions of
polarizations on the strata\/}.

\noindent Under the circumstances of Theorem \ref{kaehler1},
{\em symplectic reduction
carries a K\"ahler polarization preserved by the symmetries into a
stratified K\"ahler polarization\/}.

\noindent 2) The notion of prequantum bundle generalizes to that
of {\em stratified prequantum module\/}.
Given a complex analytic stratified K\"ahler space,
a stratified prequantum module is, roughly speaking, a system
of prequantum modules in the sense of
Definition \ref{prequantummodule}, one for the closure of each structum,
together with appropriate morphisms among them
which reflect the stratification.

\noindent 3) The notion of quantum Hilbert space
generalizes to that of {\em costratified quantum
Hilbert space\/} in such
a way that the costratified structure reflects the stratification
on the classical level.

\noindent 4) The main result says that  $ [Q,R] = 0$, that is,
quantization commutes with reduction \cite{qr}:

\begin{thm}
Under the circumstances of {\rm Theorem \ref{kaehler1}},
suppose that the K\"ahler manifold is quantizable
(that is, suppose that the cohomology class of the K\"ahler form
is integral).
When a suitable additional condition
is satisfied, reduction after quantization coincides with
quantization after reduction in the sense that not only the
reduced and unreduced quantum phase spaces correspond but the
(invariant) unreduced and reduced quantum observables as well.
\end{thm}

What is referred to here as \lq suitable additional condition\rq\
is a condition on the behaviour of the gradient flow.
For example, when the K\"ahler manifold is compact, the
condition will automatically be satisfied.

On the reduced level, the resulting classical phase
space involves in general singularities and is a
complex analytic stratified
K\"ahler space; the appropriate quantum phase space is
then a costratified Hilbert space.

In collaboration with M. Schmidt and G. Rudolph, we
plan to explore the question whether,
under certain circumstances, the
lower strata have physical significance.
This may be seen as another version of the question
spelled out above
concerning the {\em quantum structure which might have the classical
singularities as its shadow\/}.

\section{An illustration}
\label{illustration}
Let $s$ and $\ell$ be non-zero natural numbers.
The unreduced classical momentum phase space of $\ell$
particles in $\mathbb R^s$
is real affine space of real dimension $2s \ell$.
Identify this space with the vector space
$(\mathbb R^{2s})^{\times\ell}$ as usual, endow
$\mathbb R^{s}$ with the standard inner product,
$\mathbb R^{2\ell}$ with the standard symplectic structure,
and thereafter
$(\mathbb R^{2s})^{\times\ell}$ with the obvious induced
inner product and symplectic structure.
The isometry group of the inner product on
$\mathbb R^{s}$ is the orthogonal group
$\mathrm O(s,\mathbb R)$, the group of linear transformations
preserving the symplectic structure on
$\mathbb R^{2\ell}$ is the symplectic group
$\mathrm{Sp}(\ell,\mathbb R)$,
and the actions extend to
 linear $\mathrm O(s,\mathbb R)$- and
$\mathrm{Sp}(\ell,\mathbb R)$-actions
on $(\mathbb R^{2s})^{\times\ell}$ in an obvious
manner. As usual, denote the Lie algebras of
$\mathrm O(s,\mathbb R)$ and
$\mathrm{Sp}(\ell,\mathbb R)$
by
$\mathfrak {so}(s,\mathbb R)$ and
$\mathfrak{sp}(\ell,\mathbb R)$, respectively.

The $\mathrm O(s,\mathbb R)$- and
$\mathrm{Sp}(\ell,\mathbb R)$-actions
on $(\mathbb R^{2s})^{\times\ell}$
are hamiltonian.
To spell out the $\mathrm O(s,\mathbb R)$-momentum mapping
having the value zero at the origin,
identify
$\mathfrak{so}(s,\mathbb R)$ with its dual
$\mathfrak{so}(s,\mathbb R)^*$ by interpreting
 $a \in\mathfrak{so}(s,\mathbb R)$
as the linear functional on $\mathfrak{so}(s,\mathbb R)$
which assigns $\mathrm{tr}(a {}^t x)$ to
$x \in\mathfrak{so}(s,\mathbb R)$; here ${}^t x$ refers to the
transpose of the matrix $x$.
We note that, for $s \geq 3$,
\[
(s-2)\mathrm{tr}(a {}^t b) =  -\beta(a,b),\ a,b \in
\mathfrak{so}(s,\mathbb R),
\]
where $\beta$ is the {\scsc Killing\/} form of
$\mathfrak{so}(s,\mathbb R)$.
Moreover,
for a vector $\mathbf x \in \mathbb R^s$, realized as a column vector,
let ${}^t\mathbf x$ be its transpose, so that ${}^t\mathbf x$ is a row vector.
With these preparations out of the way,
the {\em angular momentum mapping\/}
\[
 \mu_{\mathrm O} \colon (\mathbb R^{2s})^{\times\ell}
 \longrightarrow \mathfrak{so}(s,\mathbb R)
\]
with reference to the origin is
given by
\[
 \mu_{\mathrm O}(\mathbf q_1,\mathbf p_1,\dots, \mathbf q_\ell,\mathbf p_\ell)
= \mathbf q_1 {}^t \mathbf p_1 -\mathbf p_1 {}^t \mathbf q_1
+\dots +\mathbf q_\ell {}^t  \mathbf p_\ell
-\mathbf p_\ell {}^t  \mathbf q_\ell.
\]
Likewise, identify
$\mathfrak{sp}(\ell,\mathbb R)$
with its dual $\mathfrak{sp}(\ell,\mathbb R)^*$
 by interpreting
 $a \in\mathfrak{sp}(\ell,\mathbb R)$
as the linear functional on $\mathfrak{sp}(\ell,\mathbb R)$ which
assigns $\frac 12 \mathrm{tr}(a x)$ to $x
\in\mathfrak{sp}(\ell,\mathbb R)$; we remind the reader that the
\emph{Killing\/} form $\beta$ of $\mathfrak{sp}(\ell,\mathbb R)$
satisfies the identity
\[
\beta(a,b) =2(\ell+1)\mathrm{tr}(ab)
\]
where $a,b \in \mathfrak{sp}(\ell,\mathbb R)$. In terms of the
present identification of $\mathfrak{sp}(\ell,\mathbb R)$ with its
dual, the $\mathrm{Sp}(\ell,\mathbb R)$-momentum mapping
\[
\mu_{\mathrm {Sp}} \colon (\mathbb R^{2s})^{\times\ell} \longrightarrow
\mathfrak{sp}(\ell,\mathbb R)
\]
having the value zero at the origin is given by the assignment to
\[
[\mathbf  q_1,\mathbf  p_1, \dots, \mathbf  q_\ell, \mathbf  p_{\ell}]
\in ( \mathbb R^s\times \mathbb R^s)^{\times \ell}
\]
of
\[\left[\begin{array}{cccc}
\left[\mathbf q_j \mathbf p_k\right] & -\left[\mathbf q_j \mathbf q_k\right]\\
\left[\mathbf p_j \mathbf p_k\right] & -\left[\mathbf p_j \mathbf q_k\right]
\end{array}
\right]
\in \mathfrak {sp}(\ell,\mathbb R),
\]
where $\left[\mathbf  q_j \mathbf  p_k\right]$ etc.
denotes the $(\ell \times \ell)$-matrix having the inner products
$\mathbf  q_j \mathbf  p_k$ etc. as entries.

Consider
the $\mathrm O(s,\mathbb R)$-reduced space
\[
N_0 = \mu_{\mathrm O}^{-1}(0)\big/ \mathrm O(s,\mathbb R).
\]
The $\mathrm{Sp}(\ell,\mathbb R)$-momentum mapping $\mu_{\mathrm {Sp}}$
induces an embedding of the reduced space
$N_0$
into $\mathfrak{sp}(\ell,\mathbb R)$.
We now explain briefly  how the image of
$N_0$ in $\mathfrak{sp}(\ell,\mathbb R)$
may be described. More details may be found in \cite{kaehler}, see also
\cite{scorza}.

Choose a positive complex structure $J$ on
$\mathbb R^{2\ell}$ which is compatible with $\omega$
in the sense that $\omega(J\mathbf u,J\mathbf v) =
\omega(\mathbf u,\mathbf v)$ for every
$\mathbf u, \mathbf v \in \mathbb R^{2\ell}$;
here \lq positive\rq\ means that the associated
real inner product $\,\cdot\,$ on
$ \mathbb R^{2\ell}$ given by
$\mathbf u \cdot \mathbf v =
\omega(\mathbf u,J\mathbf v)$
for $\mathbf u, \mathbf v \in \mathbb R^{2\ell}$
is positive definite.
The subgroup
of $\mathrm
{Sp}(\ell,\mathbb R)$
which preserves the complex structure
$J$ is a maximal compact subgroup of $\mathrm
{Sp}(\ell,\mathbb R)$;
relative to a suitable orthonormal basis,
this group comes down to a copy of the ordinary unitary group
$\mathrm U(\ell)$.
Furthermore, the
complex structure $J$ induces a {\scsc Cartan\/} decomposition
\begin{equation}
\mathfrak{sp}(\ell,\mathbb R)
 =\mathfrak{u}(\ell) \oplus \mathfrak p;
\label{symp9}
\end{equation}
here
$\mathfrak
u(\ell)$ is the Lie algebra of $\mathrm U(\ell)$,
$\mathfrak p$ decomposes as the direct sum
\[
\mathfrak p \cong \mathrm S_{\mathbb R}^2[\mathbb
R^\ell]\oplus \mathrm S_{\mathbb R}^2[\mathbb
R^\ell]
\]
of two copies of the real vector space $\mathrm S_{\mathbb R}^2[\mathbb
R^\ell]$ of real symmetric
$(\ell \times \ell)$-matrices,
and the complex structure $J$ induces
a complex structure on
$\mathrm S_{\mathbb R}^2[\mathbb
R^\ell]\oplus \mathrm S_{\mathbb R}^2[\mathbb
R^\ell]$ in such a way that
the resulting complex vector space is
complex linearly
isomorphic
to the complex vector space $\mathrm S_{\mathbb C}^2[\mathbb
C^\ell]$ of complex symmetric
$(\ell \times \ell)$-matrices
in a canonical fashion.
We refer to a nilpotent orbit $\mathcal O$ in $\mathfrak{sp}(\ell,\mathbb R)$
as being \emph{holomorphic\/}
if the orthogonal projection from $\mathfrak{sp}(\ell,\mathbb R)$
to
$\mathrm S_{\mathbb C}^2[\mathbb
C^\ell]$, restricted to $\mathcal O$, is a diffeomorphism
from $\mathcal O$ onto its image in $\mathrm S_{\mathbb C}^2[\mathbb
C^\ell]$.
The diffeomorphism from a holomorphic nilpotent orbit
$\mathcal O$ onto its image in $\mathrm S_{\mathbb C}^2[\mathbb
C^\ell]$
extends to a homeomorphism from the closure
$\overline {\mathcal O}$ onto its image in
 $\mathrm S_{\mathbb C}^2[\mathbb
C^\ell]$, and the closures of the holomorphic nilpotent orbits
constitute an ascending sequence
\begin{equation}
0 \subseteq \overline {\mathcal O}_1 \subseteq \dots \subseteq
\overline {\mathcal O}_k \subseteq \dots \subseteq \overline
{\mathcal O}_{\ell} \subseteq \mathfrak{sp}(\ell,\mathbb R),\
1 \leq k \leq \ell,
\label{symp10}
\end{equation}
such that, for $1 \leq k \leq \ell$, the orthogonal projection
from $\mathfrak{sp}(\ell,\mathbb R)$ to $\mathrm S_{\mathbb
C}^2[\mathbb C^\ell]$, restricted to $\overline{\mathcal O}_{k}$,
is a homeomorphism from $\overline{\mathcal O}_{k}$ onto the space
of complex symmetric $(\ell \times \ell)$-matrices of rank at most
equal to $k$; in particular, the orthogonal projection from
$\mathfrak{sp}(\ell,\mathbb R)$ to $\mathrm S_{\mathbb
C}^2[\mathbb C^\ell]$, restricted to $\overline{\mathcal
O}_{\ell}$, is a homeomorphism from $\overline {\mathcal
O}_{\ell}$ onto $\mathrm S_{\mathbb C}^2[\mathbb C^\ell]$.
Furthermore, each space of the kind $\overline{\mathcal O}_{k}$ is
a \emph{stratified\/} space, the stratification being given by the
decomposition according to the rank of the corresponding complex
symmetric $(\ell \times \ell)$-matrices in the homeomorphic image
in $\mathrm S_{\mathbb C}^2[\mathbb C^\ell]$.

The Lie bracket of the Lie algebra
$\mathfrak {sp}(\ell,\mathbb R)$
induces a Poisson bracket on the algebra
$C^{\infty}(\mathfrak {sp}(\ell,\mathbb R)^*)$
of smooth functions on the dual
$\mathfrak {sp}(\ell,\mathbb R)^*$
of $\mathfrak {sp}(\ell,\mathbb R)$
in a canonical fashion.
Via the identification
of $\mathfrak {sp}(\ell,\mathbb R)$ with its dual, the Lie bracket
on $\mathfrak {sp}(\ell,\mathbb R)$ induces a Poisson bracket
$\{\,\cdot\,,\,\cdot\,\}$
on  $C^{\infty}(\mathfrak {sp}(\ell,\mathbb R))$.
Indeed, the assignment to
$a \in \mathfrak {sp}(\ell,\mathbb R)$ of the linear function
\[
f_a \colon \mathfrak {sp}(\ell,\mathbb R) \longrightarrow \mathbb R
\]
given by $f_a(x) = \frac  12 \mathrm{tr}(ax)$
induces a linear isomorphism
\begin{equation}
\mathfrak {sp}(\ell,\mathbb R)
\longrightarrow
\mathfrak {sp}(\ell,\mathbb R)^*;
\label{symp11}
\end{equation}
let
\[
[ \,\cdot\, , \, \cdot \,]^* \colon
\mathfrak {sp}(\ell,\mathbb R)^*
\otimes \mathfrak {sp}(\ell,\mathbb R)^*
\longrightarrow
\mathfrak {sp}(\ell,\mathbb R)^*
\]
be the bracket on $\mathfrak {sp}(\ell,\mathbb R)^*$
induced by the Lie bracket on $\mathfrak {sp}(\ell,\mathbb R)$.
The Poisson bracket
$\{\,\cdot\,,\,\cdot\,\}$
on the algebra $C^{\infty}(\mathfrak {sp}(\ell,\mathbb R))$
is given by the formula
\[
\{f,h\}(x) = [f'(x), h'(x)]^*(x),\ x \in \mathfrak {sp}(\ell,\mathbb R).
\]
The isomorphism \eqref{symp11}
induces an embedding
of $\mathfrak {sp}(\ell,\mathbb R)$ into
$C^{\infty}(\mathfrak {sp}(\ell,\mathbb R))$,
and this embedding
is plainly a morphism
\[
\delta
\colon
\mathfrak {sp}(\ell,\mathbb R)
\longrightarrow
C^{\infty}(\mathfrak {sp}(\ell,\mathbb R))
\]
of Lie algebras when
$C^{\infty}(\mathfrak {sp}(\ell,\mathbb R))$ is viewed
as a real Lie algebra via the Poisson bracket.
In the literature,
a morphism of the kind $\delta$ is referred to as a
\emph{comomentum\/} mapping.

Let $\mathcal O$ be a holomorphic nilpotent orbit. The embedding
of the closure $\overline{\mathcal O}$ of $\mathcal O$ into
$\mathfrak {sp}(\ell,\mathbb R)$ induces a map from the algebra
$C^{\infty}(\mathfrak {sp}(\ell,\mathbb R))$ of ordinary smooth
functions on $\mathfrak {sp}(\ell,\mathbb R)$ to  the algebra
$C^{0}(\overline {\mathcal O})$ of continuous functions on
$\overline {\mathcal O}$, and we denote the image of
$C^{\infty}(\mathfrak {sp}(\ell,\mathbb R))$ in  $C^{0}(\overline
{\mathcal O})$ by $C^{\infty}(\overline {\mathcal O})$. By
construction, each function in $C^{\infty}(\overline {\mathcal
O})$ is the restriction of an ordinary smooth function on the
ambient space $\mathfrak {sp}(\ell,\mathbb R)$. Since each stratum
of $\overline {\mathcal O}$ is an ordinary smooth closed
submanifold of $\mathfrak {sp}(\ell,\mathbb R)$, the functions in
$C^{\infty}(\overline {\mathcal O})$, restricted to a stratum of
$\overline {\mathcal O}$, are ordinary smooth functions on that
stratum. Hence $C^{\infty}(\overline {\mathcal O})$ is a
\emph{smooth structure\/} on $\overline {\mathcal O}$. The algebra
$C^{\infty}(\overline {\mathcal O})$ is referred to as the algebra
of {\scsc Whitney\/}-smooth functions on $\overline {\mathcal O}$,
relative to the embedding of $\overline {\mathcal O}$ into the
affine space $\mathfrak {sp}(\ell,\mathbb R)$. Under the
identification \eqref{symp11}, the orbit $\mathcal O$ passes to a
\emph{coadjoint\/} orbit. Consequently, under the surjection
$C^{\infty}(\mathfrak {sp}(\ell,\mathbb R)) \to
C^{\infty}(\overline {\mathcal O})$, the Poisson bracket
$\{\,\cdot\,,\,\cdot\,\}$ on the algebra $C^{\infty}(\mathfrak
{sp}(\ell,\mathbb R))$ descends to a Poisson bracket on
$C^{\infty}(\overline {\mathcal O})$, which we still denote by
$\{\,\cdot\,,\,\cdot\,\}$, with a slight abuse of notation. This
Poisson algebra turns $\overline {\mathcal O}$ into a stratified
symplectic space. Combined with the complex analytic structure
coming from the projection from $\overline {\mathcal O}$ onto the
corresponding space of complex symmetric $(\ell \times
\ell)$-matrices, in this fashion, the space $\overline {\mathcal
O}$ acquires a \emph{complex analytic stratified K\"ahler space\/}
structure. The composite of the above comomentum mapping $\delta$
with the projection from $C^{\infty}(\mathfrak {sp}(\ell,\mathbb
R))$ to $C^{\infty}(\overline {\mathcal O})$ yields an embedding
\begin{equation}
\delta_{\mathcal O}
\colon
\mathfrak {sp}(\ell,\mathbb R)
\longrightarrow
C^{\infty}(\overline {\mathcal O})
\label{delta}
\end{equation}
which is still a morphism of Lie algebras
and therefore a comomentum mapping in the appropriate sense.

The $\mathrm{Sp}(\ell,\mathbb R)$-momentum mapping
induces an embedding of the reduced space
$N_0$
into $\mathfrak{sp}(\ell,\mathbb R)$ which identifies
$N_0$ with the closure
$\overline {\mathcal O}_{\min(s,\ell)}$
of the holomorphic nilpotent orbit
$\mathcal O_{\min(s,\ell)}$ in $\mathfrak{sp}(\ell,\mathbb R)$.
In this fashion, the reduced space
$N_0$ inherits a complex analytic stratified K\"ahler structure.
Since the
$\mathrm{Sp}(\ell,\mathbb R)$-momentum mapping
induces an
identification
of $N_0$ with $\overline {\mathcal O}_{s}$
for every $s \leq \ell$ in a compatible manner,
the ascending sequence \eqref{symp10},
and in particular the notion of holomorphic nilpotent orbit,
is actually independent of the choice of complex structure $J$ on
$\mathrm R^{2 \ell}$.
For a single particle, i.~e. $\ell = 1$, the description of the reduced space
$N_0$ comes down to that of the semicone given
in Section \ref{example} above.

In the case at hand, we will now explain briefly the quantization procedure
developed in \cite{qr}.
Suppose that $s \leq \ell$ (for simplicity), let $m =
s \ell$,
and endow the affine coordinate ring of $\mathbb C^{m}$,
that is, the polynomial algebra $\mathbb C[z_1,\dots,z_m]$,
with the inner product $\,\cdot\,$
given by the standard formula
\begin{equation}
\psi \cdot\psi' = \int \psi \overline{\psi'}
\mathrm e^{-\frac {\mathbf z \overline {\mathbf z}} 2} \varepsilon_m,
\quad
\varepsilon_m= \frac {\omega^m}{(2 \pi)^m m!},
\label{innerprod}
\end{equation}
where $\omega$ refers to the symplectic form on
$\mathbb C^{m}$.
Furthermore, endow
the polynomial algebra $\mathbb C[z_1,\dots,z_m]$
with the induced $\mathrm O(s,\mathbb R)$-action.
By construction, the affine complex coordinate ring
$\mathbb C[\overline {\mathcal O}_s ]$
of $\overline {\mathcal O}_s$
is canonically isomorphic to the algebra
\[
\mathbb C[z_1,\dots,z_m]^{\mathrm O(s,\mathbb R)}
\]
of $\mathrm O(s,\mathbb R)$-invariants in $\mathbb C[z_1,\dots,z_m]$.
The restriction of the inner product $\,\cdot\,$
to $\mathbb C[\overline {\mathcal O}_s ]$
turns
$\mathbb C[\overline {\mathcal O}_s ]$ into a pre-Hilbert space, and
{\scsc Hilbert\/}
space completion yields a {\scsc Hilbert\/} space
which we write as $\widehat{\mathbb C}[\overline {\mathcal O}_s ]$.
This is the Hilbert space which arises by
\emph{holomorphic quantization\/}
on the complex analytic stratified K\"ahler space
$\overline {\mathcal O}_s$;
see \cite{qr} for details.
On this Hilbert space, the elements of the Lie algebra
$\mathfrak u(\ell)$ of the unitary group $\mathrm U(\ell)$
act in an obvious fashion; indeed,
the elements of
$\mathfrak u(\ell)$, viewed as functions
in $C^{\infty}(\overline {\mathcal O}_s)$,
are classical observables which are directly quantizable,
and quantization yields the obvious
$\mathfrak u(\ell)$-representation on
$\mathbb C[\overline {\mathcal O}_s ]$.
This construction may be carried out for any $s \leq \ell$
and, for each
$s \leq \ell$,
 the resulting quantizations yields a \emph{costratified
Hilbert space\/} of the kind
\begin{equation}
\mathbb C  \longleftarrow \widehat{\mathbb C}[\overline {\mathcal
O_1}] \longleftarrow \ldots \longleftarrow \widehat {\mathbb
C}[\overline {\mathcal O_s}]. \label{costratified}
\end{equation}
Here each arrow is just a restriction mapping and is actually a
morphism of representations for the corresponding quantizable
observables, in particular, a morphism of $\mathfrak
u(\ell)$-representations; each arrow amounts essentially to an
orthogonal projection. Plainly, the costratified structure
integrates to a costratified $\mathrm U(\ell)$-representation,
i.~e. to a corresponding system of $\mathrm
U(\ell)$-representations. The resulting costratified quantum phase
space for $\overline {\mathcal O_s}$ of the kind
\eqref{costratified}  may be viewed as a {\em singular\/} Fock
space. This quantum phase space is entirely given in terms of {\em
data on the reduced level\/}.

Consider the unreduced classical harmonic oscillator
energy $E$ which is given by
$E=z_1 \overline z_1 + \dots +
z_m \overline z_m$; it quantizes to the Euler operator (quantized
harmonic oscillator hamiltonian).
For $s\leq \ell$,
the reduced classical phase space $Q_s$ of $\ell$ harmonic oscillators in
$\mathbb R^s$ with total angular momentum zero
and fixed energy value (say) $2k$ fits into an ascending sequence
\begin{equation}
 Q_1 \subseteq \dots \subseteq Q_s \subseteq
\dots \subseteq Q_{\ell} \cong \mathbb C \mathrm P^d
\label{project}
\end{equation}
of complex analytic stratified K\"ahler spaces where
\[
\mathbb C \mathrm P^d =\mathrm P(\mathrm S^2[\mathbb C^\ell]),
\quad d = \frac {\ell(\ell+1)}2-1.
\]
The sequence \eqref{project} arises from the sequence \eqref{symp10}
by \emph{projectivization\/}.
The parameter $k$ (or rather $2 k$) is
encoded in the Poisson structure. Let $\mathcal O(k)$ be the
$k$'th power of the hyperplane bundle on $\mathbb C \mathrm P^d$,
let
\[
\iota_{Q_s}\colon Q_s \longrightarrow Q_{\ell} \cong \mathbb C
\mathrm P^d
\]
be the inclusion, and let
$
\mathcal O_{Q_s}(k)= \iota_{Q_s}^*\mathcal O(k).
$
The  quantum \emph{Hilbert} space amounts now to the space of holomorphic
sections of $\iota_{Q_s}^*\mathcal O(k)$,
and the resulting {\em costratified quantum Hilbert space\/} has the form
\[
\Gamma^{\mathrm{hol}}(\mathcal O_{Q_1}(k)) \longleftarrow \ldots
\longleftarrow \Gamma^{\mathrm{hol}}(\mathcal O_{Q_s}(k)).
\]
Each vector space $\Gamma^{\mathrm{hol}}(\mathcal O_{Q_{s'}}(k))$
($1 \leq s' \leq s$) is a
finite-dimensional
representation space for the quantizable
observables in $C^{\infty}(Q_{s})$,
in particular, a $\mathfrak{u}(\ell)$-representation, and this
representation
integrates to a $\mathrm U(\ell)$-representation, and each arrow
is a morphism of representations; similarly as before,
these arrows are just restriction maps.

We will now give a description of the decomposition of the space
\[
\Gamma^{\mathrm{hol}}(\mathcal O_{Q_\ell}(k)) = S_{\mathbb C}^k
[\mathfrak p^*]
\]
of homogeneous degree $k$ polynomial functions on $\mathfrak p
=S_{\mathbb C}^2[\mathbb C^\ell]$ into its \emph{irreducible\/} $\mathrm
U(\ell)$-representations in terms of highest weight vectors.
To this end we note that
coordinates $x_1,\dots,x_\ell$ on $\mathbb C^\ell$ give
rise to coordinates of the kind $\{x_{i,j} = x_{j,i}; \, 1 \leq i,j \leq
\ell\}$ on $S_{\mathbb C}^2 [\mathbb C^{\ell}]$, and the
determinants
\[
 \delta_1 = x_{1,1}, \
\delta_2 = \left | \begin{array}{cccc} x_{1,1}& x_{1,2}\\
                           x_{1,2}& x_{2,2}
                   \end{array}\right|,\
\delta_3 = \left | \begin{array}{cccc} x_{1,1}& x_{1,2} & x_{1,3}\\
                           x_{1,2}& x_{2,2} & x_{2,3}\\
                           x_{1,3}& x_{2,3} & x_{3,3}\\
                   \end{array} \right|,
\ \text{etc.}
\]
are highest weight vectors for certain $\mathrm
U(\ell)$-re\-pre\-sen\-ta\-tions.
For $1 \leq s \leq r$
and $k \geq 1$, the $\mathrm U(\ell)$-representation
$\Gamma^{\mathrm{hol}}(\mathcal O_{Q_s}(k))$ is the sum of the
irreducible representations having as highest weight vectors the
monomials
\[
\delta_1^{\alpha} \delta_2^{\beta} \ldots \delta_s^{\gamma}, \quad
\alpha +2 \beta + \dots + s\gamma = k,
\]
and the restriction morphism
\[
 \Gamma^{\mathrm{hol}}(\mathcal O_{Q_s}(k))\longrightarrow
 \Gamma^{\mathrm{hol}}(\mathcal O_{Q_{s-1}}(k))
\]
has the span of the representations involving $\delta_s$ explicitly
as its kernel and,
restricted to the span of those irreducible representations which
do \emph{not\/} involve
$\delta_s$, this morphism is an isomorphism.

This situation may be interpreted
in the following fashion: The composite
\[
\mu_{2k}\colon \overline {\mathcal O}_s \subseteq
\mathfrak{sp}(\ell,\mathbb R)\cong\mathfrak{sp}(\ell,\mathbb R)^*
\longrightarrow
\mathfrak u(\ell)^*
\]
is a singular momentum mapping for the
$\mathrm U(\ell)$-action on
$\overline {\mathcal O}_s$;
actually, the adjoint
$\mathfrak u(\ell) \to C^{\infty}(\overline {\mathcal O}_s)$
of $\mu^{2k}$
amounts to the composite of \eqref{delta} with the inclusion of
$\mathfrak u(\ell)$ into $\mathfrak {sp}(\ell,\mathbb R)$.
The {\sl irreducible $\mathrm U(\ell)$-representations which
correspond to the coadjoint orbits in the image
\[
\mu_{2k}(O_{s'}\setminus O_{s'-1}) \subseteq \mathfrak u(\ell)^*
\]
of the stratum $O_{s'}\setminus O_{s'-1}$ ($1 \leq s' \leq s$) are
precisely the irreducible representations having as highest weight
vectors the monomials
\[
\delta_1^{\alpha} \delta_2^{\beta} \ldots \delta_{s'}^{\gamma}
\quad (\alpha +2 \beta + \dots + s'\gamma = k)
\]
involving $\delta_{s'}$ explicitly, i.~e. with\/} $\gamma \geq 1$.

\section{Applications and outlook}

In \cite{atibottw}, {\scsc Atiyah and Bott\/}
raised the issue of \emph{determining the singularities\/}
of moduli spaces of semistable holomorphic vector bundles
or, more generally, of moduli spaces of semistable principal bundles
on a non-singular complex projective curve.
The complex analytic stratified K\"ahler structure
which we isolated on a moduli space of this kind,
as explained in Example 2 above,
actually determines the singularity structure;
in particular, near any point,  the structure may be understood in terms
of a suitable local model.
The appropriate notion of singularity
is that of singularity in the sense of stratified
K\"ahler spaces; this notion depends on the entire structure,
not just on the complex analytic structure.
Indeed, the examples spelled out above (the exotic plane
with a single vertex, the exotic plane with two vertices,
the 3-dimensional complex projective space with the Kummer surface
as singular locus, etc.) show
that a point of
a stratified K\"ahler space may well be a singular point without
being a complex analytic singularity.

A number of applications
of the theory of stratified K\"ahler spaces
have already been mentioned.
Using the approach to lattice gauge theory in \cite{kan},
we intend to develop elsewhere a rigorous approach to
the quantization of certain lattice gauge theories by means of
the K\"ahler quantization scheme for complex analytic
stratified K\"ahler spaces
explained in the present paper.
In collaboration with M. Schmidt and G. Rudolph, we
plan to apply this scheme to  situations
of the kind explored in \cite{kirutwo} --\cite{kirufou}
and to compare it with the approach to quantization
in these papers.
Constrained quantum systems occur
in molecular mechanics as well, see e.~g. \cite{taniiwai}
and the references there.
Perhaps the K\"ahler quantization scheme for complex analytic
stratified K\"ahler spaces will shed new light on these quantum systems.

\section*{Acknowledgments}

This paper was written during a stay at the Institute for
theoretical physics at the University of Leipzig. This stay was
made possible by the Deutsche Forschungsgemeinschaft in the
framework of a Mercator-professorship, and I wish to express my
gratitude to the Deutsche Forschungsgemeinschaft. It is a pleasure
to acknowledge the stimulus of conversation with J. Kijowski, G.
Rudolph, and M. Schmidt.

\bigskip

\noindent Universit\'e des Sciences et Technologies de Lille, UFR
de Math\'ematiques, CNRS-UMR 8524
\\ 59655 VILLENEUVE D'ASCQ, C\'edex, France
\\ and
\\ Institute
for Theoretical Physics, Universit\"at Leipzig\\
04109 LEIPZIG, Germany \\
{Johannes.Huebschmann@math.univ-lille1.fr}
\end{document}